\documentclass[twocolumn,aps,showpacs,prb,tightenlines,amsmath,amssymb]{revtex4}
\usepackage{bm}
\usepackage{graphicx}              
\usepackage{amssymb}               
\usepackage{amsmath}                     
\usepackage{colordvi}
\usepackage{calrsfs} 
\usepackage{mathrsfs}
\usepackage[thinlines,thiklines]{easybmat}
\newcommand{\bgreek}[1]{\mbox{\boldmath$#1$\unboldmath}}
\begin{document}   

\title{Novel valley
  depolarization dynamics and valley Hall effect of exciton\\
 in mono- and bilayer MoS$_2$} 
\author{T. Yu}
\author{M. W. Wu}
\thanks{Author to whom correspondence should be addressed}
\email{mwwu@ustc.edu.cn.}
\affiliation{Hefei National Laboratory for Physical Sciences at
  Microscale,  Key Laboratory of Strongly-Coupled Quantum Matter Physics and Department of Physics, 
University of Science and Technology of China, Hefei,
  Anhui, 230026, China} 
\date{\today}

\begin{abstract} 
We investigate the valley depolarization dynamics 
 and valley Hall effect of exciton due to the electron-hole exchange
 interaction in mono- and bilayer MoS$_2$ by
  solving the kinetic spin Bloch equations. The effect of
  the exciton energy
  spectra by the electron-hole exchange interaction is explicitly considered. 
 For the valley depolarization dynamics, in the monolayer MoS$_2$, it is found
 that in the strong scattering regime, the
  conventional motional narrowing picture is no longer valid, and a novel valley
  depolarization channel is opened.
For the valley Hall effect of exciton, in both the mono- and bilayer MoS$_2$,
 with the exciton equally pumped in the K and K' valleys, the
system can evolve into the equilibrium state where the valley polarization is 
parallel to the effective magnetic field due to the exchange interaction. With
the drift of this equilibrium state by applied uniaxial strain, the
exchange interaction can induce the {\it momentum-dependent} valley/photoluminesence
polarization, which leads to the valley/photoluminesence Hall current. Specifically, the disorder
strength dependence of the valley Hall conductivity is revealed. In the strong
scattering regime, the valley Hall conductivity decreases with the increase of
the disorder strength; whereas in the weak scattering regime, it saturates to 
a constant, which can be much larger than the one in Fermi system due
to the absence of the Pauli blocking. 
\end{abstract}

\maketitle
\section{Introduction}
In recent years, as a new candidate to realize the
 valleytronics,
monolayer (ML) and bilayer (BL) transition metal dichalcogenides (TMDs) have attracted much
attention.\cite{MoS_1,MoS_2,XiaoDi,Yaowang_review,Valley_exciton,Glazov_review}  
 To efficiently control the valley degree of
 freedom in ML and BL TMDs,
 both the
 optical\cite{MoS_2,absorption_Mak,valley_wang24,valley_wang18,
absorption_wang15,Xinhuizhang,Qingmingzhang,MoSe2_Marie,WSe2_Marie,
Cui1,Cui2,Marie_BL,Spin_layer_locking,WSe2_Marie2}
 and electrical\cite{XiaoDi,Valley_Hall_vasp,Science,Science2}
 techniques have been explored. For the optical
 method, the chiral optical
valley selection rule allows for the optical creation of the valley polarization,
 which are mainly realized by the excitonic
 excitation.\cite{MoS_2,absorption_Mak,valley_wang24,
valley_wang18,absorption_wang15,Xinhuizhang,Qingmingzhang,MoSe2_Marie,
WSe2_Marie,Cui1,Cui2,Marie_BL,Spin_layer_locking,WSe2_Marie2}
 For the electrical
 method, due to the contrast Berry curvature for the electron or hole in the K
 and K' valleys, the
 valley Hall effect of electron or hole has been predicted,\cite{XiaoDi} and then 
 confirmed by the experiments in ML\cite{Science} and BL\cite{Science2} MoS$_2$.
 Furthermore, the method combining both the optical and
 electrical techniques to realize the valley Hall effect of trion is proposed
 theoretically.\cite{Dirac_cone}
 This proposal is based on the fact that
 the four configurations of the trions in ML TMDs can obtain nonzero Berry
 curvature due to the electron-electron, hole-hole and electron-hole (e-h) exchange
 interactions.\cite{Dirac_cone} 
 It can be seen that
 ML\cite{MoS_2,absorption_Mak,valley_wang24,valley_wang18,absorption_wang15,
Xinhuizhang,Qingmingzhang,MoSe2_Marie,WSe2_Marie,XiaoDi,Valley_Hall_vasp,WSe2_Marie2}
and BL\cite{Cui1,Cui2,Marie_BL,Spin_layer_locking} TMDs provide an ideal platform to study the rich
valley dynamics based on the valley polarization or valley current. Accordingly,
on one hand, it is important to study the lifetime of the valley polarization,
i.e., the valley depolarization dynamics; on the other hand, it is useful to 
 explore efficient
 methods to create and control the valley current.  

The valley depolarization dynamics in ML and BL TMDs has been extensively
  studied, showing rich 
 features for different members.\cite{MoS_2,absorption_Mak,valley_wang24,
valley_wang18,absorption_wang15,
Xinhuizhang,Qingmingzhang,MoSe2_Marie,
WSe2_Marie,WSe2_Marie2,XiaoDi,Cui1,Cui2,Marie_BL,Spin_layer_locking}
 In ML TMDs, it has been experimentally found that the steady-state
 valley polarization can be large (around 50\%) in MoS$_2$,
 WS$_2$ and WSe$_2$,\cite{MoS_2,Xinhuizhang,WSe2_Marie}
 whereas extremely
 small (around 5\%) in MoSe$_2$.\cite{Qingmingzhang,MoSe2_Marie} Fast valley
 depolarization with the lifetime about picoseconds
 due to the inter-valley exciton transition is observed.\cite{many_body,CD} It is
 theoretically shown that due to the
 strong Coulomb interaction, the e-h exchange interaction\cite{Yu_ML,Glazov,MacDonald,Polariton,Louie} can
 provide an efficient valley depolarization channel based on the
 Maialle-Silva-Sham (MSS) mechanism.\cite{Sham1,Sham2} Including the e-h
  Coulomb exchange interaction, the Hamiltonian of the exciton expressed by the
  center-of-mass momentum ${\bf k}$  is written as\cite{Yu_ML}
\begin{equation}
H_{\rm ML}=\frac{\displaystyle \hbar^2{\bf k}^2}{\displaystyle 2m_{\rm
    ex}}+Q(k)\left(\begin{array}{cc}
{\bf k}^2 & -k_{+}^2\\
-k_{-}^2 &{\bf k}^2
\end{array}\right),
\label{Exchange_ML}
\end{equation}
in which the first and second terms represent the kinetic energy and e-h exchange interaction, respectively. 
$m_{\rm ex}=m_e+m_h$ is the exciton mass with $m_e$ and
$m_h$ being the electron and hole masses; $Q(k)=\frac{\displaystyle e^2}{\displaystyle
  2\varepsilon_0\kappa (|{\bf k}|+\kappa_{\rm sc})}|\phi_{1s}^{2D}(0)|^2
\alpha_{\rm ML}$ and $k_{\pm}=k_x\pm ik_y$.
 Here, $\varepsilon_0$ and $\kappa$ stand for
 the vacuum permittivity and relative
dielectric constant; $\kappa_{\rm sc}$ is the screening
 wavevector; $\phi^{\rm 2D}_{1s}({\bf r})=\sqrt{8/{\pi
    a_B^2}}e^{-2r/a_B}$ represent the exciton ground-state wavefunction 
with ${\bf r}$ being the relative
coordinate of the electron and hole, and $a_B$ denoting the exciton radius;
 $\alpha_{\rm ML}$ is the material parameter.\cite{Yu_ML}

 This exchange interaction can cause the ``precession'' of the
 exciton states with a ${\bf k}$-dependent frequency ${\Omega}({\bf k})$,
 which causes the inhomogeneous broadening.\cite{Sham1,Sham2}
 In analogy to the D'yakonov-Perel' (DP)
 mechanism,\cite{DP} in the
 strong scattering regime with $|\Omega({\bf k})|\tau_k\ll 1$ with
 $\tau_k$ being the momentum relaxation time,
 the valley depolarization time is
 estimated to be $\tau_s^{-1}=\langle\Omega^2({\bf k})\rangle\tau_k$.
 Here $\langle\cdot\cdot\cdot\rangle$
 denotes the ensemble average.
 Accordingly, it seems that the valley
 depolarization should be always suppressed
 by the momentum
 scattering in the strong scattering regime.
 Nevertheless, Yu {\it et al.} showed that with the e-h exchange interaction,
 the energy spectra of the exciton is modified to be the Dirac
 cone.\cite{Dirac_cone} With this large modification of the exciton spectra, the
 momentum scattering should also be markedly influenced by the exchange interaction. Although
 it is then pointed out that the Dirac cone does not exit due to the existence
 of the
 intra-valley e-h exchange interaction,\cite{MacDonald,Polariton,Louie} it is
 demonstrated that the
 exchange interaction modifies the energy spectra of the
 exciton markedly.\cite{MacDonald,Polariton,Louie} This motivates us to study the valley dynamics with
 the exchange interaction explicitly modifying the energy spectra.

Similar to the ML situation, in BL TMDs, the exchange interaction between the
  four degenerate states labeled by the valley and layer
  indices is also expected to cause
  the photoluminesence (PL) depolarization.\cite{Yu_BL} However, it is experimentally found that
 different from ML situation, the steady-state PL polarization in the BL
 WS$_2$ and WSe$_2$ can be much larger than the one in ML under the same experimental
 conditions.\cite{Cui1,Cui2,Marie_BL,Spin_layer_locking} Then it is theoretically
predicted that for the BL WS$_2$, with the {\it isotropic} dielectric constant,
 there exits a steady state with the PL  polarization being always half of the
 initial one due to the specific form of the exchange interaction,
 indicating that the valley depolarization time can be very 
 long in BL TMDs.\cite{Yu_BL} Specifically, the exchange interaction Hamiltonian between
 the four degenerate {\it intra-layer} exciton states is
 written as 
\begin{equation}
H^{\rm BL}_{\rm ex}({\bf k})\approx{\tilde{Q}(k)}\left(\begin{array}{cccc}
{\bf k}^2 & \gamma k_{+}^2 & -k_{+}^2 & -\gamma {\bf k}^2\\
\gamma k_{-}^2 & {\bf k}^2 & -\gamma {\bf k}^2  & -k_{-}^2\\
-k_{-}^2&-\gamma {\bf k}^2   & {\bf k}^2  &\gamma k_{-}^2\\
-\gamma {\bf k}^2 & -k_{+}^2  & \gamma k_{+}^2  & {\bf k}^2\\
\end{array}\right).
\end{equation} 
Here, $\tilde{Q}(k)=\frac{\displaystyle e^2}{\displaystyle
  2\varepsilon_0\kappa_{\parallel} (|{\bf k}|+\kappa_{\rm sc})}|\tilde{\phi}_{1s}^{2D}(0)|^2
\tilde{\alpha}_{\rm BL}$ with $\kappa_{\parallel}$ denoting the intra-layer relative
dielectric constant; $\gamma=\sqrt{\kappa_{\parallel}/\kappa_{\perp}}$ 
with $\kappa_{\perp}$ being the inter-layer relative
dielectric constant. The tilde labels that the parameters in BL situation can be
different from the ones in ML. Similar to the ML situation,
 this exchange interaction can also markedly modify the
energy spectra of the exciton, which is also expected to influence the valley
dynamics. Moreover, the study for the PL depolarization dynamics for the {\it anisotropic} dielectric
constant\cite{dielectric_BL} is still lacking.

 From above analysis, it can be seen  that by treating ``valley'' as ``spin'' in
 ML TMDs, 
 the exchange interaction actually plays the role of the
 spin-orbit coupling (SOC)
 in the electronic system. In the electronic system,
 the intrinsic spin Hall effect has been well understood in
 the system with the SOC in the weak scattering limit, which is absent for the
 Rashba but can exist for other type of the SOC.\cite{SHE_MacDonald,SHE_KSBE,SHE_Glazov,SHE_Ka,cancellation_0,
cancellation_1,cancellation_2,cancellation_3,cancellation_4,cancellation_5}
 Moreover, it is found that the intrinsic spin Hall conductivity is a constant
 in the clean sample, which is nevertheless less
studied in the strong scattering regime.\cite{SHE_MacDonald,SHE_KSBE,SHE_Glazov,SHE_Ka,cancellation_0,
cancellation_1,cancellation_2,cancellation_3,cancellation_4,cancellation_5}
 Then it is natural to expect that in analogy to the
 {\it intrinsic} spin Hall effect of electrons,\cite{SHE_MacDonald,SHE_KSBE,SHE_Glazov,SHE_Ka,cancellation_0,
cancellation_1,cancellation_2,cancellation_3,cancellation_4,cancellation_5}
 there exists the ``valley'' Hall effect of exciton due to the
 exchange interaction in ML TMDs. Accordingly,
 with the generation of the exciton current, which can be realized by applying the uniaxial
 strain,\cite{Nagaosa} the valley current perpendicular to the exciton
 current can emerge in ML TMDs. 
 However, in the BL TMDs, the exchange interaction exists between
{\it four}
rather than two 
degenerate exciton states, which is very different from the electronic
system with two degenerate spin bands. It is an
interesting problem to study whether there exists the valley Hall
effect of exciton for the four-state system. 
It is emphasized that in the previous works, the ``spin'' Hall effect
of exciton has been
proposed.\cite{exciton_Hall_SOC,exciton_Hall_Berry}
  However, it is different from the
 proposal here. In the work of Wang {\it et al.},\cite{exciton_Hall_SOC} the
 ``spin'' Hall effect arises due to the different strength of the SOC experienced
 by the electron and hole in the exciton; whereas in the work of Yao {\it et
   al.},\cite{exciton_Hall_Berry} it arises from the Berry curvature, which is in analogy to the
 intrinsic 
 anomalous Hall effect of electron.\cite{anomalous_Hall_1,anomalous_Hall_2}

 In the present work, by explicitly
  considering exciton energy
  spectra modified by the e-h exchange interaction, we investigate the valley
  depolarization dynamics and valley Hall effect of exciton in ML and BL MoS$_2$ by
  solving the kinetic spin Bloch equations (KSBEs).\cite{wu-review}
 For the valley depolarization dynamics, in the ML MoS$_2$, it is
 found that with the exchange-interaction-modified energy spectra, in the
  strong scattering regime, the
  conventional relation $\tau_s\propto \tau_k^{-1}$ is no longer valid. It is
  shown that a novel valley
  depolarization channel is opened in the {\it strong} scattering regime,
 where the valley lifetime first decreases and
  then increases with the increase of the disorder strength, showing
  the Elliott-Yafet\cite{Yafet,Elliott} (EY) like behavior in the DP mechanism from the
  point of view of the spin
  relaxation.\cite{Awschalom,Zutic,fabian565,Dyakonov,Korn,wu-review}
  This channel comes from the inhomogeneous broadening from the
    module of the momentum of the exciton in the exciton-disorder
    scattering, in which the same
    energy corresponds to different momentum module with the exchange-interaction-modified
    energy spectra. This is very different from the conventional
    situation, in which the inhomogeneous broadening comes from
    the angular anisotropy of the momentum in the exciton-disorder
    scattering.\cite{Sham1,Sham2} Moreover, due to the enhancement of the module-dependent inhomogeneous broadening
 by the momentum scattering, it shows EY-like behavior in the MSS
  mechanism.
 For the BL MoS$_2$, the PL depolarization dynamics with both the isotropic and
  anisotropic dielectric constants is investigated. With the isotropic
  dielectric constant, it is shown that with the exchange interaction modifying
  the energy spectra, the steady state revealed in our previous
  work\cite{Yu_BL}
 still exists. Whereas with the anisotropic dielectric
constant, the steady
state vanishes. However, it is found that when the dielectric constant 
is close to the isotropic situation, the PL polarization first
decreases fast and then slowly, indicating that the effective
depolarization time can also be much longer than the
ML situation. 

For the valley Hall effect of exciton, the valley Hall conductivity for the
ML and BL MoS$_2$ in both the weak and strong scattering regimes are calculated by
the KSBEs. It
is shown that with the exciton in the K and K' valleys equally pumped, the
system
 evolves into the equilibrium state where the valley polarization is 
parallel to the effective magnetic field due to the exchange interaction. With
the drift of this equilibrium state due to the applied uniaxial strain, this parallelism is broken and hence the
effective magnetic field can induce the {\it momentum-dependent} out-of-plane valley/PL
polarization, which accounts for the valley/PL current of exciton. Furthermore, the disorder
strength dependence of the valley Hall conductivity is revealed. For both the ML
and BL MoS$_2$, in the strong
scattering regime, the valley Hall conductivity decreases ($\propto \tau_k^2$) with the increase of
the disorder strength; whereas in the weak scattering regime, the valley Hall
conductivity saturates to a constant. Specifically, it is found that the
valley Hall conductivity in the weak scattering regime is proportional to the
population of the exciton with zero momentum. By further considering that with the Bose
distribution (therefore no the Pauli blocking), this population can
be extremely large at low temperature and high exciton density. Accordingly, the valley Hall
conductivity for the exciton can be much larger than the one for the Fermi
system.
 All these behaviors can be well understood analytically
 in the weak
exchange interaction limit.
 
This paper is organized as follows. In Sec.~{\ref{Model}}, we set
up the model and KSBEs. In
Sec.~{\ref{Monolayer}},
 we study the valley depolarization dynamics and valley Hall
 effect of exciton in ML MoS$_2$. Specifically, in Sec.~{\ref{Novel_channel}}, a novel valley
   depolarization channel is presented; in Sec.~{\ref{valley_ML}, the disorder
     strength dependence 
     of the valley
     Hall effect of exciton is studied first numerically and then understood
     analytically. 
 In Sec.~{\ref{Bilayer}}, the valley depolarization dynamics and valley Hall
 effect of exciton are further discussed in BL MoS$_2$. We
summarize in Sec.~{\ref{summary}}.

\section{Model and KSBEs}
\label{Model}

We start the investigation from the set up of the kinetic equation for the exciton by
  considering the exciton-disorder scattering. We
  first present the effective
  Hamiltonian for the exciton-disorder interaction expressed by the 
  center-of-mass coordinate of the exciton, which is derived in
  Refs.~\onlinecite{Disorder1,Disorder2}. 
The Hamiltonian in the disordered system is written as
\begin{eqnarray}
\nonumber
\hspace{-0.3cm}&&\Big[-\frac{\hbar^2}{2m_e}\nabla^2_{e}-\frac{\hbar^2}{2m_h}\nabla^2_{h}
-\frac{e^2}{4\pi\varepsilon_0\kappa|{\bf
  r}_e-{\bf r}_h|}+W_e({\bf r}_e)\\
\hspace{-0.3cm}&&\mbox{}+W_h({\bf
r}_h)\Big]\Psi_{\eta}({\bf r}_e,{\bf r}_h)=E_{\eta}\Psi_{\eta}({\bf r}_e,{\bf
r}_h),
\label{disorder_Hamiltonian}
\end{eqnarray}
where $W_e({\bf r}_e)$ and $W_h({\bf r}_h)$ denote the {\it intra-valley}
disorder potential for the electron and hole, respectively, 
 and $\eta$ labels the exciton state including the valley and layer indices.
 For ML MoS$_2$, we do not consider
 the inter-valley scattering, which is suppressed
 for the hole with large splitting of the valence band,
 because the spin-flip scattering is forbidden unless the mirror reflection 
 symmetry is broken.\cite{WangLin1,WangLin2,YangFei,Yu_ML}
 For BL MoS$_2$, the inter-layer scattering is further neglected
 because only the hole with the same spin can hop between different
 layers, which is nevertheless very weak due to the large splitting of the valencen
 bands, and the inter-layer hopping for the electron is forbidden due
 to the lattice symmetry.\cite{Cui1,Cui2,Marie_BL,Spin_layer_locking,Yu_BL}
When the disorder is not very strong, which does not influence the relative motion
    of the exciton, the exciton-disorder interaction can be treated
    perturbatively and expressed by the
    center-of-mass coordinate.\cite{Disorder1,Disorder2} By focusing on the ground
    state ($1s$-state), the center-of-mass
    part of the Hamiltonian reads  
\begin{equation}
\Big[-\frac{\hbar^2\nabla^2_{\bf R}}{2m_{\rm ex}}+V_{\rm ex}({\bf R})\Big]\Psi_{1s}({\bf
  R})=E_{1s}\Psi_{1s}({\bf R}),
\label{center-of-mass}
\end{equation}
where ${\bf R}=(m_e{\bf r}_e+m_h{\bf r}_h)/m_{\rm ex}$.
Here, with $m_e\approx m_h$ in ML and BL MoS$_2$, 
\begin{eqnarray}
\nonumber
\hspace{-0.08cm}V_{\rm ex}({\bf R})\approx 4\int d{\bf R}'
  \Big|\phi^{\rm 2D}_{1s}\big(2{\bf R}'-2{\bf R}\big)\Big|^2\big[W_e({\bf
    R}')+W_h({\bf
    R}')\big]\\
\hspace{-0.08cm}
\label{potential}
\end{eqnarray}
describes the effective exciton-disorder interaction. From
Eq.~(\ref{potential}), one notices that the charged impurity is inefficient for
the exciton-disorder interaction.

Furthermore, with the e-h exchange interaction Hamiltonian $H_{\rm ex}({\bf k})$ included
in Eq.~(\ref{center-of-mass}), the exciton dynamics under the uniaxial strain 
 can be described by the KSBEs including the coherent, drift and scattering terms:\cite{wu-review}
\begin{equation}
  \partial_t\rho_{\bf k}=\partial_t\rho_{\bf
      k}|_{\rm coh}+\partial_t\rho_{{\bf
    k}}|_{\rm drift}+\partial_t\rho_{\bf k}|_{\rm  scat}.
\label{ksbe}
\end{equation}
In these equations, $\rho_{\bf k}$  represent the $n\times n$ density matrices of
exciton with the center-of-mass momentum ${\bf k}$ at time $t$,
 in which the diagonal terms describe the exciton distribution
 functions and off-diagonal terms
represent the inter-state coherence. Specifically, $n=2$ and 4 for the ML and BL
MoS$_2$, respectively.

In the
  {\it collinear} space, the coherent term is given by 
\begin{equation}
\partial_t\rho_{\bf k}|_{\rm
   coh}=-(i/\hbar)\big[H_{\rm ex}({\bf k}),\rho_{\bf
  k}\big],
\end{equation}
where $[\ ,\ ]$ stands for the commutator.
The drift term is denoted as
\begin{equation}
\partial_t\rho_{\bf
    k}|_{\rm drift}=-({\bf F}/\hbar)\cdot\nabla_{\bf k}\rho_{\bf
    k},
\end{equation}
where ${\bf F}$ represents the external force field due to the applied uniaxial strain. Finally, 
the scattering term
 $\partial_t\rho({\bf k},t)|_{\rm  scat}$ due to the exciton-disorder
 scattering is written as
\begin{eqnarray}
\nonumber
&&\partial_t\rho_{\bf k}|_{\rm
   scat}=-\frac{\pi}{\hbar}\sum_{{\bf
    k}'\eta_1\eta_2}|U_{\bf k-k'}|^2\delta(E_{{\bf
    k}',\eta_1}-E_{{\bf k},\eta_2})\\
&&\mbox{}\times\Big[\big(T_{{\bf k}',\eta_1}T_{{\bf
    k},\eta_2}\rho_{{\bf k}}-T_{{\bf k},\eta_2}T_{{\bf k}',\eta_1}\rho_{{\bf
    k}'}\big)+{\rm H.c.}\Big].
\label{KSBEs}
\end{eqnarray} 
In Eq.~(\ref{KSBEs})
\begin{eqnarray}
\nonumber
|U_{\bf q}|^2&=&\int\int d{\bf r}d{\bf r'}\langle[U({\bf
  r})-U_0] [U({\bf r'})-U_0]\rangle e^{-i{\bf q}\cdot({\bf
    r}-{\bf r'})}\\
&=&\int\int d{\bf r}d{\bf r'}C({\bf r}-{\bf r'})e^{-i{\bf q}\cdot({\bf
    r}-{\bf r'})}\equiv C_{\bf q},
\label{UUq}
\end{eqnarray}
with $U_0$ being the average value of the disorder potential. 
$C({\bf q})$ is taken to be the Gaussion correlation function\cite{Disorder1,Disorder2} 
\begin{equation}
\nonumber
C({\bf q})=\pi V_R^2\sigma_R^2\exp(-\sigma_R^2 q^2/4),
\label{Gaussion}
\end{equation}
where $V_R$ is the potential amplitude and $\sigma_R$ denotes the radius of
  the correlation length of the disorder. Specifically, when $\sigma_R q\ll 1$,
  Eq.~(\ref{Gaussion}) actually describes the short-range exciton-disorder
  interaction. $E_{{\bf k},\eta}$ and $T_{{\bf
      k},\eta}$ are the energy spectra of the exciton
    and the projection matrix, whose expressions are given explicitly in Appendix~\ref{AA} for both the ML
    and BL situations.

\section{Monolayer MoS$_2$}
\label{Monolayer}

In this section, we investigate the valley depolarization dynamics and valley Hall
  effect for the {\it A}-exciton
 in the ML MoS$_2$. All parameters including the band
structure and material parameters used in our computation are
listed in Table~\ref{material_ML}.
\begin{table}[htb]
  \caption{Parameters used in the computation for ML MoS$_2$.}
  \label{material_ML} 
  \begin{tabular}{l l l l}
    \hline
    \hline
    $m_e/m_0$&\;\;\;\;\;$0.35^{a}$&\;\;\;\;\;\;\;\;\;$T$ (K) &\;\;\quad 20\\
    $m_h/m_0$&\;\;\;\;\;$0.44^{a}$&\;\;\;\;\;\;\;\;\;$n_{\rm ex}$ (cm$^{-2}$)&\;\;\quad$10^{11}$\\
    $\kappa$&\;\;\;\;\;$3.43^a$&\;\;\;\;\;\;\;\;\;$n_{\rm e}$
    (cm$^{-2}$)&\;\;\quad$1.6\times 10^{10}$\\
    $a_B$(nm)&\;\;\quad$3.0$&\;\;\;\;\;\;\;\;\;$\sigma_R$ (nm)&\;\;\;\quad$6$\\
    $\alpha_{\rm ML}$ (${\rm \AA}^2$)&\;\;\quad$4.91^{b}$&\;\;\;\;\;\;\;\;\;$V_{R0}$
    (meV)&\;\;\quad$4.4\times 10^{-2}$\\
    $F$ (${\rm eV/cm}$)&\;\;\quad$10$&\\
    \hline
    \hline
\end{tabular}\\
 \;\;
\hspace{-5.5cm}$^a$Reference [\onlinecite{directgap_wang23}].\\
\hspace{-5.5cm}$^b$Reference [\onlinecite{Yu_ML}].
\end{table}

In Table \ref{material_ML}, $T$ denotes the temperature, which is low enough that the
exciton-phonon interaction is
neglected;\cite{Disorder1,Disorder2,WangLin1,WangLin2,YangFei}
 $n_{\rm ex}$ and $n_e$ are the
exciton and electron densities, with the former much larger than the
latter. With these parameters, $\kappa_{sc}=\frac{\displaystyle
  e^2m_e}{\displaystyle 2\hbar^2\varepsilon_0\kappa_0}\frac{\displaystyle
  1}{\displaystyle e^{-\mu_e/(k_BT)}+1}\approx 1.5\times 10^{9}$/m, with
$\mu_e$ being the chemical potential of electron.

\subsection{Novel Valley Depolarization Dynamics}
\label{Novel_channel}
In this subsection, we investigate the valley depolarization dynamics,
  especially focus on a novel valley depolarization channel in ML MoS$_2$.
 The valley depolarization time is obtained by
  solving the KSBEs from the temporal evolution of the valley polarization
$P(t)=\sum_{\bf k}{\bf S}_{\bf k}^z(t)/n_{\rm ex}
=\sum_{\bf k}\mbox{Tr}[\rho_{\bf k}(t) \hat{\sigma}_z]/n_{\rm ex}$, with $\hat{\sigma}_z$
being the $\hat{z}$-component of the Pauli matrix. 
According to the chiral optical
valley selection rule,\cite{MoS_2,absorption_Mak}
 by using the elliptically polarized light, the system is initialized to be
\begin{equation}
\rho_{\bf k}(0)=\frac{B_{{\bf k}\uparrow}+B_{{\bf k}\downarrow}}{2}
+\frac{B_{{\bf k}\uparrow}-B_{{\bf k}\downarrow}}{2}{\hat{\sigma}_z}.
\label{polarization}
\end{equation}
In Eq.~(\ref{polarization}), $B_{{\bf k}\sigma}=\{\exp[(\varepsilon_{{\bf
    k}}-\mu_{\sigma})/(k_BT)]-1\}^{-1}$
 is the Bose-Einstein distribution function at temperature $T$,
 with $\varepsilon_{\bf k}=\hbar^2k^2/(2m_{\rm ex})$ and 
 $\mu_{\uparrow,\downarrow}$
 standing for the chemical potentials determined by the exciton 
density $n_{\rm ex}$=$\sum_{\bf k}$Tr[${\rho_{\bf k}}$] and the initial
valley polarization $P(0)$. $P(0)=10\%$ in our numerical calculation.

\subsubsection{Analytical Analysis on the Conventional Situation}

For comparison with the novel valley depolarization channel addressed in the next
  subsection (Sec.\ref{Novel_results}),
 we first present the analytical analysis of the conventional MSS 
 mechanism,\cite{Sham1,Sham2,DP} which has been used to understand the recent
 experimental results.\cite{WSe2_Marie2,Xinhuizhang} 
As we know, when the splitting energy due to the exchange interaction, which
is referred
to as the ``exchange energy'' in this work, is much {\it smaller} than the kinetic
 energy $\varepsilon_{\bf k}$, the exchange energy can be neglected in the energy spectra,
 i.e., $E_{{\bf
    k},\eta}\approx \varepsilon_{\bf k}$ in
Eq.~(\ref{KSBEs}).
 Based on this approximation, Eq.~(\ref{KSBEs}) becomes 
\begin{eqnarray}
\partial_t\rho_{\bf k}|_{\rm
   scat}\approx -\frac{2\pi}{\hbar}\sum_{{\bf
    k}'}|U_{\bf k-k'}|^2\delta(\varepsilon_{{\bf
    k}'}-\varepsilon_{\bf k})\big(\rho_{{\bf k}}-\rho_{{\bf
    k}'}\big).
\label{KSBEs_app}
\end{eqnarray}
To find the valley depolarization time, we transform the KSBEs from the collinear
space to the helix one by the unitary transformation $\tilde{\rho}_{\bf
  k}=U_{\bf k}^{\dagger}\rho_{\bf k}U_{\bf k}$.\cite{JinLuo} In
the helix representation, the KSBEs become 
\begin{eqnarray}
\nonumber
\hspace{-0.3cm}&&\partial_t \tilde{\rho}_{\bf k}+i(\varepsilon_k^{\rm
ex}/\hbar)[\hat{\sigma}_z,\tilde{\rho}_{\bf k}]
+(2\pi/\hbar) a^2\sum_{\bf k'}\delta(\varepsilon_{{\bf k}'}-\varepsilon_{\bf
k})\\
\hspace{-0.3cm}&&\mbox{}\times\big(\tilde{\rho}_{\bf k}-S_{\bf kk'}\tilde{\rho}_{\bf k'} S_{\bf
  k'k}\big)=0,
\label{helix}
\end{eqnarray}
where the exchange energy $\varepsilon_k^{\rm
ex}=Q(k)k^2$ and $S_{\bf kk'}=U_{\bf k}^{\dagger}U_{\bf k'}$.
 Here, for simplicity, we consider the situation with $\sigma_R
 |{\bf k}-{\bf k'}|\ll 1$, hence $|U_{\bf k-k'}|^2$ is replaced by the
 constant $a^2=\pi V_R^2\sigma_R^2$
 in Eq.~(\ref{helix}).
After Fourier analysis with $\tilde{\rho}_{\bf k}=\sum_l\tilde{\rho}_k^le^{il\theta_{\bf k}}$,
one finds that the zeroth order of the density matrix $\tilde{\rho}_k^0$ forms a
closed equation, 
 \begin{equation}
\frac{\partial \tilde{\rho}_k^0}{\partial t}+\frac{i}{\hbar}\varepsilon_k^{\rm
  ex}[\hat{\sigma}_z,\tilde{\rho}_k^0]
+\frac{\tilde{\rho}_k^0}{2\tau}-\frac{1}{2\tau}\hat{\sigma}_x\tilde{\rho}_k^0\hat{\sigma}_x=0,
\label{closed}
\end{equation}
which is not obvious in the collinear representation.\cite{Zhang_graphene}
Here, $1/\tau=m_{\rm ex}a^2/\hbar^3$ is the momentum scattering rate.

By further noticing that ${\bf \tilde{S}}_{\bf k}^x={\bf S}_{\bf k}^z$, one obtains from
Eq.~(\ref{closed}) that 
\begin{eqnarray}
\nonumber
\hspace{-0.6cm}&&{\bf S}_{\bf k}^z(t)=\frac{P(0)}{2}\Big(1+\frac{1}{\sqrt{1-16\Omega_k^2\tau^2}}\Big)
e^{\big(-\frac{t}{2\tau}+\frac{t}{2\tau}\sqrt{1-16\Omega_k^2\tau^2}\big)}\\
\hspace{-0.6cm}&&\mbox{}+\frac{P(0)}{2}\Big(1-\frac{1}{\sqrt{1-16\Omega_k^2\tau^2}}\Big)
e^{\big(-\frac{t}{2\tau}-\frac{t}{2\tau}\sqrt{1-16\Omega_k^2\tau^2}\big)},
\label{Solution_conventional}
\end{eqnarray}
where $\Omega_k=\varepsilon_k^{\rm ex}/\hbar$ is the precession frequency
between different exciton states.
From Eq.~(\ref{Solution_conventional}), it is obtained that in the strong
scattering limit with $\Omega_k\tau\ll 1$, 
\begin{equation}
{\bf S}_{\bf k}^z(t)\approx P(0)\exp(-4\Omega_k^2\tau t),
\label{strong}
\end{equation}
and hence the valley depolarization time $\tau_v\approx [4\langle\Omega_k^2\rangle\tau]^{-1}$
 is inversely proportional to the
momentum scattering
time, which is the motional narrowing effect in the random-walk
 theory.\cite{Awschalom,Zutic,fabian565,Dyakonov,Korn,wu-review} 
Whereas in the weak scattering limit with $\Omega_k\tau\gg 1$,
\begin{equation}
{\bf S}_{\bf k}^z(t)\approx P(0)e^{-t/(2\tau)}\cos(2\Omega_kt).
\label{weak_scattering}
\end{equation}
Hence, two factors influence the valley depolarization in the weak scattering
limit. On one hand, the momentum scattering opens a valley depolarization
channel due to the factor $e^{-t/(2\tau)}$; on the other hand, the factor
$\cos(2\Omega_kt)$ can cause free induction decay due to different precession
 frequency with different momentum (inhomogeneous broadening).\cite{wu-review,Zhang_graphene}

However, with the {\it same} initial state [Eq.~(\ref{polarization})],
 above conventional picture obtained by 
 the weak exchange energy approximation
is no longer
valid when the exchange energy is comparable to or even larger than the kinetic
one,\cite{Yu_ML,MacDonald,Polariton,Louie} which is shown in
the next subsection.

\subsubsection{Momentum Scattering Dependence of the Novel Valley Depolarization}
\label{Novel_results}

In ML MoS$_2$, due to the strong Coulomb interaction and the large exciton mass,
the exchange energy is comparable to the kinetic
one.\cite{Yu_ML,MacDonald,Polariton,Louie}
 This is
true even when the screening effect due to the residue electron in the sample
 is considered (Table~\ref{material_ML}).
Therefore, the exchange energy should enter the energy spectra in the
scattering term when calculating the 
valley depolarization time. Consequently, a novel valley
  depolarization channel in the strong scattering regime, in which the valley
  depolarization is enhanced rather than suppressed by the
  momentum scattering, showing the EY-like
 behavior,\cite{Yafet,Elliott} is switched on.  

To understand the new valley depolarization channel, 
 we focus on a simplified model, where only the diagonal elements
 in the projection matrix [Eq.~(\ref{projection_ML})] are retained.
 The corresponding scattering term
reads
\begin{equation}
\partial_t\rho_{\bf k}|_{\rm scat}\approx -\frac{1}{4\tau}\sum_{\eta_1\eta_2}\int
d\varepsilon_{k'}\delta(E_{{\bf k'},{\eta_1}}-E_{{\bf
  k},{\eta_2}})(\rho_{\bf k}-\rho_{k'}^0),
\label{channel}
\end{equation} 
where ${\rho}_{k'}^0=\frac{1}{2\pi}\int d\theta_{\bf k'}\rho_{\bf
  k'}$. Furthermore, by means of the Fourier analysis, the density matrix is expanded as
 $\rho_{\bf k}=\rho_k^0+\sum_{l\ne 0}\rho_{\bf k}^le^{il\theta_{\bf k}}$, where
 the zeroth and non-zeroth components are written separately. It can be seen that for the zeroth order
of $\rho_{\bf k}$, the scattering term 
\begin{equation}
\partial_t\rho_{\bf k}|_{\rm scat}\approx -\frac{1}{4\tau}\sum_{\eta_1\eta_2}\int
d\varepsilon_{k'}\delta(E_{{\bf k'},{\eta_1}}-E_{{\bf
  k},{\eta_2}})(\rho_{k}^0-\rho_{k'}^0)
\label{channel2}
\end{equation} 
is nonzero for the exciton-disorder scattering between different energy
branches; whereas it is forbidden in the conventional
situation from Eq.~(\ref{KSBEs_app}).

 Therefore, Eq.~(\ref{channel2})
opens an additional valley depolarization channel by causing the
module-dependent inhomogeneous broadening. This is very different from the conventional
situation, in which the inhomogeneous broadening
arises from the angular anisotropy of the momentum in
the exciton-disorder scattering.\cite{Sham1,Sham2} With this enhancement
  of the inhomogeneous broadening, the valley
  depolarization tends to be enhanced. However, this additional
  channel also enhances the momentum
  scattering, which tends to suppress the valley depolarization. Therefore, there exists the
  competition between the effective inhomogeneous
  broadening and momentum scattering in this new channel. It
  is demonstrated that the EY-like behavior exactly comes from this enhancement of
  the inhomogeneous broadening. It is shown in
  Fig.~\ref{figyw1} that compared to the full calcualtion (the red
  solid curve with circles), when the
  additional valley depolarization channel is removed, 
the EY-like behavior vanishes in the pink
chain curve.

\begin{figure}[ht]
  {\includegraphics[width=8.15cm]{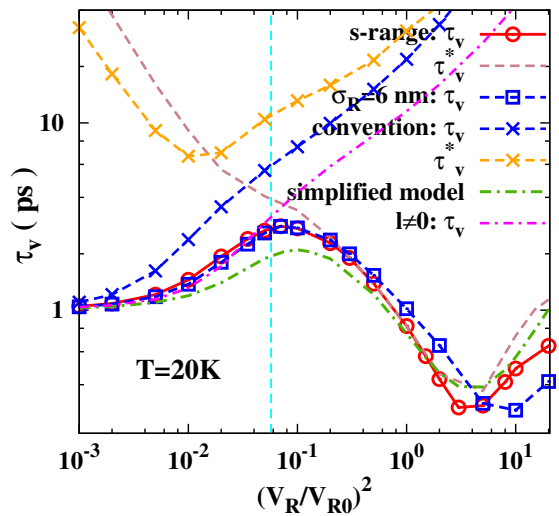}}
  \caption{(Color online) Disorder strength dependence of the valley
    depolarization time in ML MoS$_2$. The boundary between the weak and strong
    scattering regimes [$2\langle\Omega_k\rangle\tau\approx 1$]
 is denoted by the vertical cyan dashed line. The valley depolarization times 
    with both the short-range 
    scattering (the red solid curve with circles) and Gaussion correlation
    function with $\sigma_R=6$~nm (the blue dashed curve with squares) 
     are shown. With
    short-range scattering, the red solid curve with circles and
    gray dashed curve correspond to the valley depolarization times fitted from
    $\sum_{\bf k}{\bf S}_{\bf k}$ and $\sum_{\bf k}|{\bf S}_{\bf k}|$,
    respectively.
    For comparison, the conventional situation is plotted by the blue
    ($\sum_{\bf k}{\bf S}_{\bf k}$) and
    orange ($\sum_{\bf
      k}|{\bf S}_{\bf k}|$) dashed curve  with crosses. Finally, the green chain curve corresponds to
 the simplified model [Eq.~(\ref{channel})], and the pink chain curve
 represents the situation with Eq.~(\ref{channel2}) removed from Eq.~(\ref{KSBEs}).}
  \label{figyw1}
\end{figure}

Figure~\ref{figyw1} shows the disorder strength dependence of the valley
depolarization time computed, based on the material parameters shown in
Table~\ref{material_ML}. In Fig.~\ref{figyw1}, the boundary between the weak
and strong scattering regimes [$(V_R/V_{R0})^2\approx 0.06$] is shown
as the vertical cyan dashed line.
Accordingly, it is shown that with both the short-range (the red solid
curve with circles) and Gaussian correlation funciton (the blue
dashed curve with squares) in the scattering term,  
 in the weak
scattering regime, the valley depolarization time increases with the increase of
the disorder strength; whereas in the strong scattering regime, the valley
depolarization is first enhanced and then suppressed by the momentum
scattering, which is referred to as the EY-like and normal strong
scattering regimes. Furthermore, these two curves
for the short-range and Gaussion correlation function coincide with each other, 
showing that the short-range scattering is a good approximation for the
exiton-disorder scattering here. This is because for the Bose-Einstein
distribution at low temperature here, $q\sigma_R\ll 1$ is satisfied,
 and hence the exponential function in Eq.~(\ref{Gaussion}) can be neglected.

For comparison, the conventional situation is also computed, shown by the blue dashed
  curve with crosses in Fig.~\ref{figyw1}.
 It can be seen that with the increase of the disorder strength, the valley
  depolarization time increases monotonically in both the weak and strong
  scattering regimes. In the strong scattering regime, this confirms
  Eq.~(\ref{strong}) where the valley depolarization is suppressed by the
  momentum scattering. Whereas in the weak scattering regime, from
  Eq.~(\ref{weak_scattering}), it seems that the
 valley depolarization should be enhanced by the momentum
  scattering. However, because the energy dispersion in
 the Bose-Einstein distribution, the free
  induction decay dominates the valley depolarization, which can be suppressed
  by the momentum scattering. To see this point, we plot the valley
  depolarization time $\tau_v^*$ fitted from incoherently
 summed spin polarization $\sum_{\bf
    k}|{\bf S}_{\bf k}|$,\cite{Kuhn,Haug,Wu_Metiu,Wu_Ning}
 where the free
  induction decay is destroyed. It is shown by the orange dashed curve with
  crosses that with the
  increase of the disorder strength, $\tau_v^*$ decreases in the weak
  scattering regime because the
momentum scattering can directly open a channel for the valley depolarization
[$e^{-t/(2\tau)}$ in Eq.~(\ref{weak_scattering})], but increases in the strong scattering regime.

 It is interesting to see that the behavior of the valley depolarization with the
 exchange-interaction-modified energy spectra is similar to the
 conventional situation in the weak scattering regime, 
but very different in the strong scattering regime.
In the weak scattering regime, the new valley polarization
  channel is not important, which can be seen from Fig.~\ref{figyw1} that the valley
  depolarization with (green chain curve) and without (pink chain
  curve)
 this channel almost coincides with each other. Therefore, same as the conventional
situation, due to the suppression of the free induction decay by the scattering,
the valley depolarization time increases with the increase of the disorder
strength. By further fitting $\tau_v^*$ from incoherently summed spin
polarization 
 $\sum_{\bf k}|{\bf S}_{\bf k}|$ (gray dashed curve), one observes $\tau_v^*$ decreases
with the increase of the disorder strength in the weak scattering regime due to
the destroy of the free induction decay. In the EY-like regime, the enhancement of the valley
depolarization by the scattering comes from the enhancement of inhomogeneous
broadening due to the novel valley
depolarization channel [Eq.~(\ref{channel2})]. Finally, in the normal strong
scattering regime, the momentum scattering is very strong. The enhancement of
the momentum scattering in Eq.~(\ref{channel2}) becomes more important than the
enhancement of the inhomogeneous broadening, and hence the valley depolarization
is suppressed by the momentum scattering.

Finally, it can be seen from Fig.~\ref{figyw1} that the green chain curve
calculated from the simplified model [Eq.~(\ref{channel})] almost coincides with the one by full
calculation (the red solid curve with circles). Therefore, it seems that the off-diagonal elements in the
   projection matrix play less important role in the valley depolarization in the regimes
 we study. However, it influences the behavior of the temporal evolution of the valley
 polarization in the normal strong scattering regime, leading to the
 oscillations of the valley polarization (refer to Appendix~\ref{BB}).

\subsection{Valley Hall Effect of Exciton}
\label{valley_ML}
In this part, we study the valley Hall effect of exciton in ML MoS$_2$ both numerically and
analytically. In the calculation, the initial state is set to be the equally populated Bose-Einstein
distribution in the K and K' valleys, i.e., 
\begin{equation}
\rho_{\bf k}(0)=B^0_{k}\hat{I},
\label{nopolarization}
\end{equation}
where $B^0_{k}$ is the Bose-Einstein distribution function.
The computation parameters are list in Table~\ref{material_ML}.
With this experimentally-realized initial state,\cite{MoS_2,absorption_Mak}
 we first show that the system can evolve to the
equilibrium state in which the ``spin'' vectors [${\bf S}^x_{\bf
  k}=\mbox{Tr}(\rho_{\bf k}\hat{\sigma}_x)$, ${\bf S}_{\bf k}^y=\mbox{Tr}(\rho_{\bf k}\hat{\sigma}_y)$ and
${\bf S}_{\bf k}^z$] are parallel to the ${\bf
  k}$-dependent magnetic field due to the exchange interaction [Sec.~\ref{equlibrium_state}].
 Then we show that with this equilibrium state, after applying the
external force field due to the uniaxial
strain,\cite{Nagaosa,exciton_Hall_Berry}
 the drift of this equilibrium state can induce the valley
Hall current of exciton (Sec.~\ref{Valley_Hall_ML}). 

\subsubsection{Equilibrium State without External Force Field}
\label{equlibrium_state}

Before the concrete study of the valley Hall effect of exciton in ML MoS$_2$, it
is important to know the property of
the equilibrium state with
the exciton initially equally-populated in the K and K' valleys.
In Figs.~\ref{figyw2}(a) and (b), it is shown that after a long time, the system evolves into
the equilibrium state which corresponds to the ``spin'' separation for ${\bf
  S}_{\bf k}^x$ and ${\bf S}_{\bf k}^y$
in the momentum space, respectively.
 Specifically, in this equilibrium state, the spin vectors ${\bf S}_{\bf k}^x$
 and ${\bf S}_{\bf k}^y$ are parallel to the ${\bf
  k}$-dependent magnetic field due to the exchange interaction
 ${\bgreek \Omega}_{\bf k}^x=-2\varepsilon_k^{\rm ex}\cos(2\theta_{\bf k})$ and ${\bgreek
  \Omega}_{\bf k}^y=-2\varepsilon_k^{\rm ex}\sin(2\theta_{\bf k})$. Below we analytically demonstrate this property
in the weak exchange interaction limit.
\begin{figure}[ht]
  {\includegraphics[width=7.9cm]{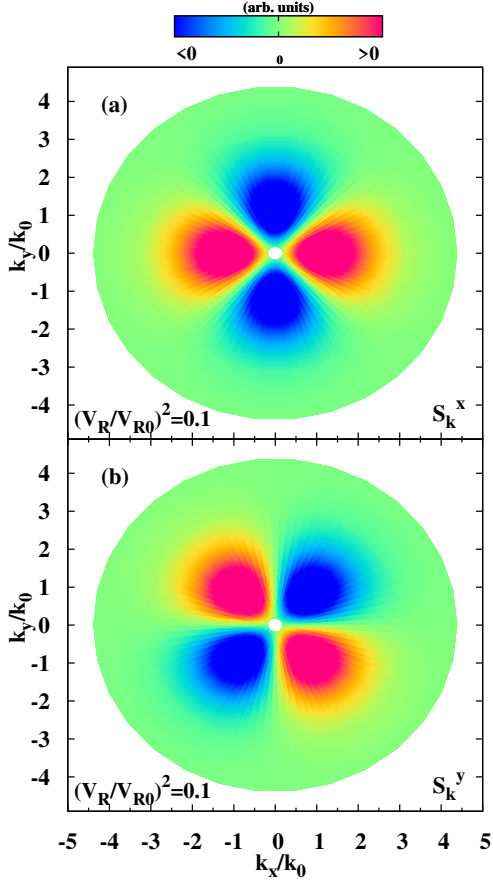}}
  \caption{(Color online) Momentum distribution of the ``spin'' vectors ${\bf
      S}_{\bf k}^x$ (a)
    and ${\bf S}_{\bf k}^y$ (b) in the equilibrium state. $k_0\approx 7.9\times
    10^7$/m is the Fermi wave-vector of the system.
 The calculation shows that in the equilibrium state, ${\bf S}_{\bf
      k}^x\propto \cos(2\theta_{\bf k})$ and ${\bf S}_{\bf
      k}^y\propto \sin(2\theta_{\bf k})$, which are parallel to the in-plane effective
    magnetic field due to the exchange interaction along $\hat{x}$-direction [${\bgreek \Omega}_{\bf
      k}^x=-2\varepsilon_k^{\rm ex}\cos(2\theta_{\bf k})$] and $\hat{y}$-direction [${\bgreek
  \Omega}_{\bf k}^y=-2\varepsilon_k^{\rm ex}\sin(2\theta_{\bf k})$], respectively.}
  \label{figyw2}
\end{figure}

In the weak exchange interaction limit, we expand the energy spectra in the
  linear order of the exchange energy in the scattering term
  [Eq.~(\ref{KSBEs})],
 and then derive the equilibrium state to
be (the derivation is referred to Appendix~\ref{CC})
\begin{equation}
\rho_{\bf k}^e\approx B_k^0\hat{I}+\mathcal{H}_{\bf k}^{\rm
      ex}{\partial B_k^0}/{\partial \varepsilon_k}.
\label{steady_noE}
\end{equation}
Here, with the diagonal elements contributing to the energy spectra, the
exchange interaction Hamiltonian 
\begin{equation}
\mathcal{H}_{\bf k}^{\rm ex}=\varepsilon_k^{\rm ex}\left(\begin{array}{cc}
0 & -e^{2i\theta_{\bf k}}\\
-e^{-2i\theta_{\bf k}} & 0
\end{array}\right)
\label{Effective_SOC}
\end{equation} 
 only contains the off-diagonal elements.
This equilibrium state corresponds to the spin vectors 
 \begin{eqnarray}
&&S_{\bf k}^x=-2\varepsilon_{k}^{\rm ex}\cos(2\theta_{\bf k}){\partial
  B_k^0}/{\partial \varepsilon_k},\\
&&S_{\bf k}^y=-2\varepsilon_{k}^{\rm ex}\sin(2\theta_{\bf k}){\partial B_k^0}/{\partial \varepsilon_k}.
\end{eqnarray}  
Obviously, the spin vectors $S_{\bf k}^x$ and $S_{\bf k}^y$ are parallel to the ${\bf
  k}$-dependent magnetic field ${\bgreek \Omega}_{\bf k}^x$ and ${\bgreek \Omega}_{\bf k}^y$, respectively.

\subsubsection{Valley Hall Effect of Exciton}
\label{Valley_Hall_ML}

In this subsection, we study the valley Hall effect of exciton.
 Specifically, we explicitly show the disorder strength (momentum scattering) dependence of
the valley Hall conductivity $\sigma_{x}^z$. From the initial state
Eq.~(\ref{nopolarization}), we numerically calculate the steady state density matrix with the
applied field after long temporal evolution by the KSBEs. Then the valley Hall conductivity is calculated. 
With the valley Hall current defined as 
\begin{equation}
j^{z}_x=\sum_{\bf k}\mbox{Tr}\Big[\rho_{\bf
  k}\frac{1}{2}(\hat{\sigma}_z\hat{v}_y
+\hat{v}_y\hat{\sigma}_z)\Big]=\sigma_x^{z}F/|e|,
\end{equation}
the valley Hall conductivity $\sigma_x^{z}$ is expressed as
\begin{equation}
\sigma_x^{z}=\frac{|e|}{2F}\sum_{\bf k}\mbox{Tr}\Big[\rho_{\bf
  k}(\hat{\sigma}_z\hat{v}_y+\hat{v}_y\hat{\sigma}_z)\Big].
\label{sigma_ML}
\end{equation}
Here, $|e|$ is
the electron charge and $\hat{v}_y={\hbar
  k_y}/{m_{\rm ex}}+{\partial H_{\rm ex}^{\rm ML}({\bf k})}/{\partial k_y}$ is
the velocity operator.

 The results are shown in Fig.~\ref{figyw3}. With the weak external force field, the system is
in the linear regime, as shown in Fig.~\ref{figyw3} that 
the momentum scattering time reveals linear dependence on the disorder
strength. In this regime, it can be
  seen from Fig.~\ref{figyw3} that in the strong scattering regime, the valley Hall
conductivity decreases with the increase of the disorder strength, showing the
dependence $\sigma_{x}^z\propto \tau^2$; whereas in the weak scattering regime, the valley Hall
conductivity saturates to a constant. Below we show analytically that these features in the
momentum scattering dependence can be well understood in the weak exchange interaction
approximation.  

\begin{figure}[ht]
  {\includegraphics[width=8.1cm]{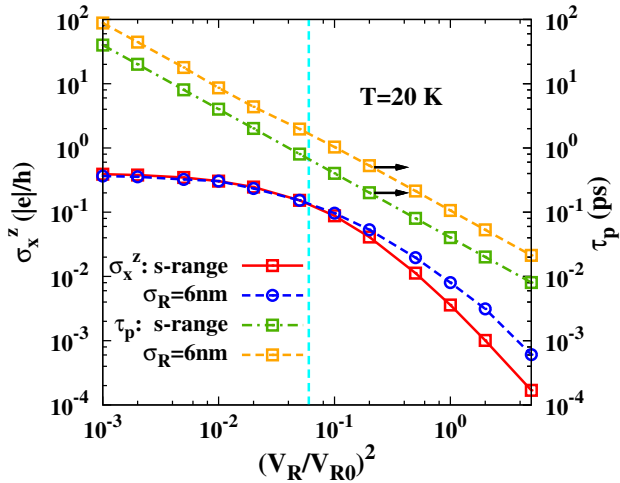}}
  \caption{(Color online) Disorder strength dependence of the valley Hall
    conductivity and the momentum scattering time (note the
scale is on the right hand side of the frame). The cyan dashed line labels the boundary between the weak and
    strong scattering regimes. The green chain (short-range)
and orange dashed (Gaussian correlation)
curves with squares show that the momentum scattering time is linearly dependent on the disorder
strength. For the valley Hall conductivity, it can be seen from the red solid
curve with squares (short-range)
and blue dashed curve with circles (Gaussian correlation) that in the strong scattering limit, the valley Hall
conductivity decreases with the increase of the disorder strength with the
dependence $\sigma_{x}^z\propto \tau^2$; whereas in the weak scattering regime, the valley Hall
conductivity saturates to a constant.}
  \label{figyw3}
\end{figure}

Here, we outline the main
 results to obtain the physical picture of the valley Hall effect of exciton.
 It has been shown that without
 the external force field, the density matrix in the equilibrium state is written as
 Eq.~(\ref{steady_noE}), which commutes with the exchange interaction
 Hamiltonian. After applying the external force field, the
   density matrix in the steady state is derived in Appendix~\ref{DD},
 as shown by Eq.~(\ref{final}).
 It is shown that based on the equilibrium state [Eq.~(\ref{steady_noE})], after applying the 
 external field, the drift part of the density matrix is [Eq.~(\ref{first_order})]
\begin{equation}
\rho_{\bf k}^{(1)}=
-\frac{F\tau}{\hbar} \frac{\partial}{\partial k_x}\big(B_k^0\hat{I}+\mathcal{H}_{\bf
k}^{\rm ex}\frac{\partial B_k^0}{\partial \varepsilon_k}\big).
\end{equation}
Obviously, this drift density matrix no longer commutes with the exchange
interaction Hamiltonian, i.e., it
can induce the momentum-dependent ``spin'' polarization along the 
$\hat{z}$-direction (valley polarization).
 This can be seen as follows. In Eq.~(\ref{final}), the
 induced density matrix responsible for 
the valley polarization is written as 
\begin{equation}
\rho_{\bf k}^{\rm in}\approx 
\frac{i}{\hbar^2}F
\frac{\partial B_k^0}{\partial \varepsilon_k}\frac{\tau^2}{1+4\Omega_k^2\tau^2}\big[\mathcal{H}_{\bf
k}^{\rm ex},\frac{\partial \mathcal{H}_{\bf k}^{\rm
ex}}{\partial k_x}\big].
\label{responsible}
\end{equation}
 Then with the exchange interaction Hamiltonian
  [Eq.~(\ref{Effective_SOC})], it can be obtained that  
\begin{eqnarray}
&&{\partial \mathcal{H}_{\bf k}^{\rm
  ex}}/{\partial k_x}\approx -2Qk_x\hat{\sigma}_x+2Qk_y\hat{\sigma}_y;\\
&&[\mathcal{H}_{\bf k}^{\rm ex},{\partial \mathcal{H}_{\bf k}^{\rm ex}}/{\partial k_x}]\approx 4iQ^2k^2k_y\hat{\sigma}_z.
\label{induction}
\end{eqnarray}
In
the derivation, we have used the fact that
 when the screening effect due to the residue electron is considered, $Q(k)$ is approximately
  a constant with our computation parameters (Table~\ref{material_ML}).
From Eq.~(\ref{induction}), one observes that the induced density matrix is
proportional to $\hat{\sigma}_z$ and dependent on the momentum $k_y$.

Then the valley Hall conductivity can be calculated. 
With the definition of the valley Hall conductivity [Eq.~(\ref{sigma_ML})],
 only the third term on the right-hand side of Eq.~(\ref{final}), i.e.,
Eq.~(\ref{responsible}), contributes to the
valley 
Hall conductivity. From Eq.~(\ref{sigma_ML}), one has
\begin{equation}
\sigma_x^{z}=-\frac{\displaystyle |e|}{\displaystyle h}\int_0^{\infty} d\varepsilon_k 
\frac{\displaystyle dB_k^0}{\displaystyle d\varepsilon_k}\frac{\displaystyle 4\Omega_k^2\tau^2}{\displaystyle
  1+4\Omega_k^2\tau^2}.
\label{sigma_ML2}
\end{equation}
From above equation, one finds that in the strong scattering regime, with
$\Omega_k\tau\ll 1$, 
\begin{equation}
\sigma_x^{z}\approx-\frac{\displaystyle 4|e|}{\displaystyle h}\int_0^{\infty} d\varepsilon_k 
\frac{\displaystyle dB_k^0}{\displaystyle d\varepsilon_k}\Omega_k^2\tau^2,
\end{equation}
which is proportional to $\tau^2$.
Whereas in the weak scattering regime with $\Omega_k\tau\gtrsim 1$,
\begin{equation}
\sigma_x^{z}\approx\frac{\displaystyle |e|}{\displaystyle h} B^0_{k=0},
\label{weak_limit}
\end{equation}
which is independent on the exchange interaction strength and momentum
scattering. Specifically, one observes that $\sigma_x^{z}$ is proportional to
$B^0_{k=0}$, which can be extremely large when the system is close to the
Bose-Einstein condensation.  
Here, with the computation parameters (Table~\ref{material_ML}), $B^0_{k=0}\approx 0.6$, and
hence $\sigma_x^{z}\approx 0.6|e|/h$, which gives a good estimate to the calculated one
$0.4|e|/h$ with strong exchange interaction. Moreover, from
Eq.~(\ref{weak_limit}), one observes that with the higher exciton
density and/or lower temperature, $B^0_{k=0}$ is large and hence the valley Hall
conductivity. According to our calculation, with the exciton density $n_{\rm
  ex}=5\times 10^{11}$~cm$^{-2}$ at 10~K, $\sigma_x^z\approx 5.5|e|/h$ in the
weak scattering limit. This is much larger than the one in Fermi system,
with the latter being limited by the Pauli blocking.\cite{SHE_KSBE,SHE_Glazov} 

Finally, we summarize the physical picture of the valley Hall effect of exciton as follows.
First of all, it is understood that in the equilibrium state, the ``spin'' vector of
 any momentum ${\bf k}$ is parallel to the ${\bf k}$-dependent effective magnetic field due
to the e-h exchange interaction (Sec.~\ref{equlibrium_state}). 
Then by applying the force field due to the uniaxial strain,\cite{Nagaosa}
 the ``spin'' vector is no longer
parallel to the effective magnetic field. Accordingly, the ``spin'' vector can
rotate around the effective magnetic field, and the {\it momentum-dependent} valley
polarization is induced. Specifically, for the exciton with
 opposite $k_y$, the
 $\hat{x}$-component of the effective magnetic field 
 is along the opposite direction, and hence the
induced valley polarization is also opposite.
 Consequently, the valley current perpendicular to the
driven exciton current is established.

 It is emphasized that the physical picture addressed above is in analogy to the intrinsic
spin Hall effect of the
electron.\cite{SHE_KSBE,SHE_Glazov,SHE_Ka,SHE_MacDonald} Nevertheless,
 two new features in this mechanism are further revealed here. On one hand, it is
 revealed that in the dirty sample corresponding to the strong
 scattering regime, the intrinsic ``spin'' Hall effect is markedly suppressed by
 the momentum scattering, with its conductivity
 proportional to $\tau^2$. On the other hand, in the weak scattering
 regime, the Bose system with no
 Pauli blocking provides
 an ideal platform to realize large ``spin'' Hall conductivity, which can
 be much larger than the one in the Fermi system, especially when the system is close to
 the Bose-Einstein condensation.

\section{Bilayer MoS$_2$}
\label{Bilayer}

In this section, we investigate the valley depolarization dynamics and valley Hall
  effect for the {\it A}-exciton, which is four-fold--degenerate, in BL
  MoS$_2$.
 In our previous work, a steady state
   in the PL depolarization dynamics in BL WS$_2$ with the isotropic dielectric
   constant was revealed in the situation without the energy
   spectra modified by the exchange interaction.\cite{Yu_BL} However, with the
   exchange-interaction-modified energy spectra, as revealed in 
   ML MoS$_2$ (Sec.~\ref{Monolayer}), the valley dynamics
   becomes very different from the conventional situation and the valley Hall
   effect of exciton can arise.
 So far, the PL depolarization dynamics for BL TMDs with anisotropic
   dielectric constant is still lacking. These motivate us to
   calculate the related PL dynamics for the four-fold--degenerate states in BL system with the energy spectra
   modified by the exchange interaction. All parameters including the band
structure and material parameters used in our computation are
listed in Table~\ref{material_BL}.
\begin{table}[htb]
  \caption{Parameters used in the computation for BL MoS$_2$.}
  \label{material_BL} 
  \begin{tabular}{l l l l}
    \hline
    \hline
    $m_e/m_0$&\;\;\;\;\;$0.35^{a}$&\;\;\;\;\;\;\;\;\;$T$ (K) &\;\;\quad 20\\
    $m_h/m_0$&\;\;\;\;\;$0.44^{a}$&\;\;\;\;\;\;\;\;\;$n_{\rm ex}$ (cm$^{-2}$)&\;\;\quad$10^{11}$\\
    $\kappa_{\parallel}$&\;\;\;\;\;$4.8^b$&\;\;\;\;\;\;\;\;\;$n_{\rm e}$
    (cm$^{-2}$)&\;\;\quad$3.3\times 10^{10}$\\
    $a_B$(nm)&\;\;\quad$3.0$&\;\;\;\;\;\;\;\;\;$\sigma_R$ (nm)&\;\;\;\quad$6$\\
    $\alpha_{\rm BL}$ (${\rm \AA}^2$)&\;\;\quad$4.51^{c}$&\;\;\;\;\;\;\;\;\;$V_{R0}$
    (meV)&\;\;\quad$5.4\times 10^{-2}$\\
    $F$ (${\rm eV/cm}$)&\;\;\quad$10$&\\
    \hline
    \hline
\end{tabular}\\
 \;\;
\hspace{-5.5cm}$^a$Reference [\onlinecite{directgap_wang23}].\\
\hspace{-5.5cm}$^b$Reference [\onlinecite{dielectric_BL}].\\
\hspace{-5.5cm}$^c$Reference [\onlinecite{Yu_BL}].
\end{table}

It is emphasized that in BL TMDs, the centro-inversion symmetry
 exists.\cite{Cui1,Cui2,Marie_BL,Spin_layer_locking,Science,Science2} Hence, 
no valley polarization but the PL polarization can be created by the chiral optical
valley selection rule.\cite{Cui1,Cui2,Marie_BL,Spin_layer_locking} Similarly,
 no valley current but the PL current can be created by the valley Hall effect due to
the symmetry, which can be measured at
 the edges of the
device channel in the experiment.\cite{Science,Science2} 

\subsection{PL Depolarization Dynamics}
\label{PL_depolarization}}
In this part, we focus on the new feature of the PL depolarization
dynamics arising in BL system compared to the ML situation. The PL depolarization dynamics is obtained by
  solving the KSBEs from the temporal evolution of the PL polarization
$P(t)=\sum_{\bf k}\mbox{Tr}[\rho_{\bf k}(t) \hat{I}_z]/n_{ex}$ with 
\begin{equation}
\hat{I}_z=\left(\begin{array}{cccc}
1 & 0 & 0 & 0\\
0 & -1 & 0  & 0 \\
0 & 0  & -1  & 0\\
0 & 0  & 0  & 1\\
\end{array}\right).
\end{equation} 
In the calculation, the initial condition is set to be
\begin{equation}
\rho_{\bf k}(0)=\frac{B_{{\bf k}\uparrow}+B_{{\bf k}\downarrow}}{2}
+\frac{B_{{\bf k}\uparrow}-B_{{\bf k}\downarrow}}{2}{\hat{I}_z}.
\label{polarization_BL}
\end{equation}
 $P(0)$ is set to be $10\%$ in our numerical calculation. Below, the PL
depolarization dynamics with isotropic and anisotropic dielectric constants are
investigated, respectively.

\begin{figure}[ht]
  {\includegraphics[width=7.9cm]{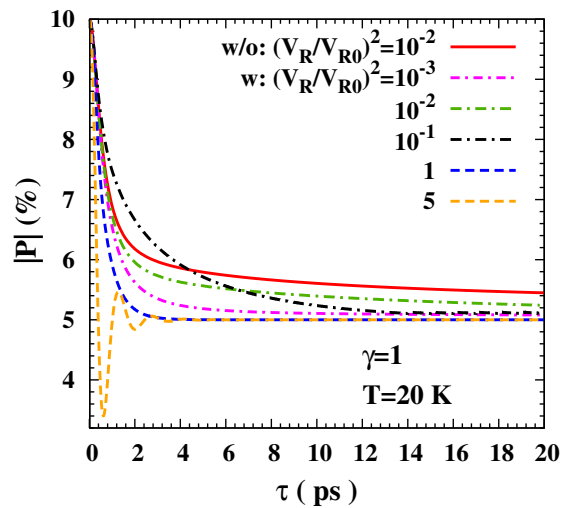}}
  \caption{(Color online) Temporal evolution of the PL polarization in the BL MoS$_2$
    with isotropic dielectric constant (${\gamma}=1$) and different disorder
    strength.
 It is shown that no matter the disorder
    strength is weak or strong, there is always
a steady state with the PL polarization being half of the initial one. The
  red solid curve is calculated without the
  exchange-interaction-modified energy spectra.\cite{Yu_BL}}
  \label{figyw4}
\end{figure}

For the isotropic dielectric constant (${\gamma}=1$), it is shown in Fig.~\ref{figyw4} that no
matter the system lies in the weak or strong scattering regime, there is always
a steady state with the PL polarization being half of the initial one. This is
the same as our previous prediction without
 the exchange-interaction-modified energy spectra (the red solid curve in Fig.~\ref{figyw4}).\cite{Yu_BL}
However, when the exchange interaction markedly modifies the energy
spectra, the density matrix in the steady state is found to be different from
the previous one.\cite{Yu_BL} 

In our previous work, with the initial condition
Eq.~(\ref{polarization_BL}), when the system
lies in the steady state, the density matrix is found to be\cite{Yu_BL}  
\begin{equation}
\rho_{\bf k}^{s}=\frac{B_{{\bf k}\uparrow}+B_{{\bf k}\downarrow}}{2}
+\frac{B_{{\bf k}\uparrow}-B_{{\bf k}\downarrow}}{4}\left(\begin{array}{cccc}
1 & 0 & 0 & 1\\
0 & -1 & -1 & 0 \\
0 & -1  & -1 & 0\\
1 & 0  & 0  & 1\\
\end{array}\right).
\label{initial3} 
\end{equation}
Here, with the exchange interaction markedly modifies the energy spectra,
 the form of the density matrix in the steady state is different, in
 which no zero elements arise and hence all states are correlated to
 each other. This can be understood in the weak exchange interaction approximation. As a simplified
model,
 with the diagonal and off-diagonal elements
entering the energy spectra of exciton, the effective exchange interaction in BL
MoS$_2$ is written as
\begin{equation}
\mathcal{\tilde{H}}^{\rm ex}_{{\bf k}}\approx{\tilde{\varepsilon}^{\rm ex}_{k}}\left(\begin{array}{cccc}
0 & \gamma e^{2i\theta_{\bf k}} & -e^{2i\theta_{\bf k}} & 0\\
\gamma e^{-2i\theta_{\bf k}} & 0 & 0  & -e^{-2i\theta_{\bf k}}\\
-e^{-2i\theta_{\bf k}}&0   & 0  &\gamma e^{-2i\theta_{\bf k}}\\
0& -e^{2i\theta_{\bf k}}  & \gamma e^{2i\theta_{\bf k}}  & 0\\
\end{array}\right),
\label{effective_SOC_BL}
\end{equation}
where $\tilde{\varepsilon}^{\rm ex}_{k}=\tilde{Q}(k)k^2$.
In the weak exchange interaction approximation,
 the KSBEs for the BL MoS$_2$ are similar to the ML
 situation (Appendices~\ref{CC} and \ref{DD}), 
\begin{eqnarray}
\nonumber
\hspace{-0.4cm}&&\frac{\partial \rho_{\bf k}}{\partial t}+\frac{F}{\hbar}\frac{\partial \rho_{{\bf k}}}{\partial
  k_x}+\frac{i}{\hbar}\big[\mathcal{\tilde{H}}_{{\bf k}}^{\rm ex},\rho_{\bf
  k}\big]+\frac{\rho_{{\bf
    k}}-\rho_{k}^0}{\tau}-\frac{\pi}{\tau}\int\frac{d\theta_{\bf
    k'}}{(2\pi)^2}\\
\hspace{-0.4cm}&&\mbox{}\times d\delta(\varepsilon_{k'}-\varepsilon_k) \{\mathcal{\tilde{H}}_{{\bf k}}^{\rm
  ex}-\mathcal{\tilde{H}}_{{\bf k'}}^{\rm ex},\rho_{\bf k}-\rho_{\bf k'}\}=0,
\label{KSBE_BL}
\end{eqnarray}
where $\{ , \}$ denotes the anti-commutator. Without the applied field, in the
steady state, the first three terms in the left-hand side of Eq.~(\ref{KSBE_BL}) are
zero. Based on the conventional density matrix in the steady state [Eq.~(\ref{initial3})]
 and following the iteration technique introduced in Appendix~\ref{CC}, one
obtains the steady state here ($\gamma=1$),
\begin{eqnarray}
\nonumber
\tilde{\rho}_{\bf k}^s&\approx& \rho_{\bf k}^s+(1/2)\{\mathcal{\tilde{H}}_{\bf k}^{\rm
      ex},{\partial\rho_{\bf k}^s}/{\partial \varepsilon_k}\}\\
&=&\rho_{\bf k}^s+(1/2)\mathcal{\tilde{H}}_{\bf k}^{\rm
      ex}({\partial B_{{\bf k}\uparrow}}/{\partial \varepsilon_{\bf k}}
+{\partial B_{{\bf k}\downarrow}}/{\partial \varepsilon_{\bf k}}).
\label{steady_noE_BL_new}
\end{eqnarray}
Obviously, Eq.~(\ref{steady_noE_BL_new}) commutes with the exchange interaction
Hamiltonian Eq.~(\ref{effective_SOC_BL}). Specifically, due to $\mathcal{\tilde{H}}_{\bf k}^{\rm
      ex}$ in Eq.~(\ref{steady_noE_BL_new}), all states become correlated to each other.

\begin{figure}[ht]
  {\includegraphics[width=7.85cm]{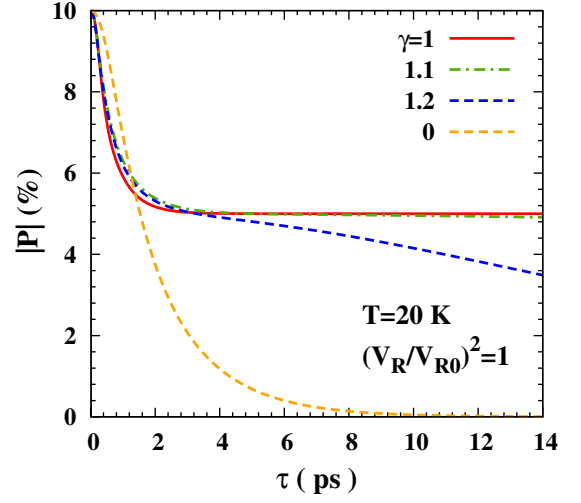}}
  \caption{(Color online) Temporal evolution of the PL polarization in the BL MoS$_2$
    with anisotropic dielectric constant. When the dielectric constant is tuned to
be anisotropic with ${\gamma}=1.1$ (the blue chain curve) and 1.2 (the
green dashed curve), the steady state vanishes. For comparison, the time
evolution of the valley polarization in ML MoS$_2$ is also plotted, which is shown as
 ${\gamma=0}$ by the orange dashed curve.}
  \label{figyw5}
\end{figure}

For the anisotropic dielectric constant, it is seen from Fig.~\ref{figyw5}
that compared to the isotropic case ${\gamma}=1$, when the dielectric constant is tuned to
be anisotropic with ${\gamma}=1.1$ (the blue chain curve) and 1.2 (the
green dashed curve), the steady state vanishes.
 However, when $\gamma$ is close to
the isotropic situation, the PL
polarization first decreases fast
and then slowly. Accordingly, the effective
depolarization time can also
be much longer than the
ML situation, shown as ${\gamma=0}$ by the orange dashed curve.

\subsection{Valley Hall Effect of Exciton}
\label{PL_Hall}
\begin{figure}[ht]
  {\includegraphics[width=8.1cm]{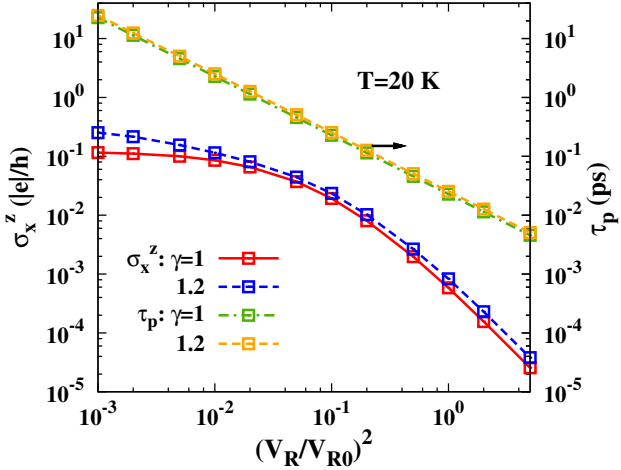}}
  \caption{(Color online) Disorder strength dependence of the valley
Hall conductivity in BL MoS$_2$. The blue chain ($\gamma=1$) and orange
dashed ($\gamma=1.2$) 
curves with squares represent the momentum
scattering time, showing the system lies in the linear regime with the applied
field (note the
scale is on the right hand side of the frame).
 It is shown that no matter the dielectric constant is isotropic with
$\gamma=1$
 (the red solid curve with squares) or
anisotropic with $\gamma=1.2$ (the blue dashed
curve with squares), in the strong scattering regime, the valley Hall
conductivity decreases with the increase of the disorder strength, showing the dependence
$\tilde{\sigma}_x^z\propto \tau^2$; whereas in the weak scattering regime, the valley
Hall conductivity saturates to a constant.}
  \label{figyw6}
\end{figure}

In this subsection, we investigate the valley Hall effect of exciton in BL
MoS$_2$. In the calculation, the initial state is set to be the equally populated Bose-Einstein
distribution in the K and K' valleys in both the upper and lower layers, i.e., 
\begin{equation}
\rho_{\bf k}(0)=B^0_{k}\hat{I}_{4\times 4}.
\label{nopolarization_BL}
\end{equation}
From the KSBEs, with the applied force field, the steady-state density matrix
is calculated and then used to calculate the valley Hall conductivity.
In analogy to the ML situation, with the PL current which carries the PL
polarization defined as 
\begin{equation}
\tilde{j}^{z}_x=\sum_{\bf k}\mbox{Tr}\Big[\rho_{\bf
  k}\frac{1}{2}(\hat{I}_z\hat{v}_y
+\hat{v}_y\hat{I}_z)\Big]=\tilde{\sigma}_x^{z}F/|e|,
\end{equation}
 the valley Hall conductivity $\tilde{\sigma}_x^{z}$ in BL MoS$_2$ is expressed as
\begin{equation}
\tilde{\sigma}_x^{z}=\frac{|e|}{2F}\sum_{\bf k}\mbox{Tr}\Big[\rho_{\bf k}
(\hat{I}_z\hat{v}_y+\hat{v}_y\hat{I}_z)\Big].
\label{sigma_BL}
\end{equation}
Here, $\hat{v}_y=\hbar k_y/\tilde{m}_{\rm ex}+{\partial H_{\rm
    ex}^{\rm BL}({\bf k})}/{\partial k_y}$.
The calculated results with both the isotropic and anisotropic dielectric
constants are summarized in Fig.~\ref{figyw6}.

In Fig.~\ref{figyw6}, the disorder strength dependence of the valley
Hall conductivity and momentum scattering time is plotted.
The curves for the momentum scattering time show that the system lies
in the linear regime.
 It can be seen from Fig.~\ref{figyw6} that in this regime, no matter the dielectric constant
is isotropic with $\gamma=1$ (the red solid
curve with squares) or
anisotropic with $\gamma=1.2$ (the blue dashed
curve with squares), in the strong scattering regime, the valley Hall
conductivity decreases with the increase of the disorder strength, showing the
dependence 
$\tilde{\sigma}_x^z\propto \tau^2$; whereas in the weak scattering regime, the valley
Hall conductivity saturates to a constant. This shows that although the
PL depolarization dynamics in the BL MoS$_2$ is different from the ML
situation, the properties of the valley Hall conductivity between them are
similar. Below, the valley Hall conductivity is also derived in the weak
exchange interaction approximation, with the steady-state density matrix with the applied
field derived in Appendix~\ref{DD}.

As a simplified model, the exchange interaction Hamiltonian
Eq.~(\ref{effective_SOC_BL}) is used.
 From Eq.~(\ref{sigma_BL}), it can be seen that only the third term on
 the left-hand side of
Eq.~(\ref{final}) contributes to the
valley 
Hall conductivity. In Eq.~(\ref{final}), with the exchange interaction Eq.~(\ref{effective_SOC_BL}),
 one finds
\begin{eqnarray}
\nonumber
\hspace{-0.5cm}&&\Big[\mathcal{\tilde{H}}_{{\bf k}}^{\rm ex},{\partial \mathcal{\tilde{H}}_{{\bf k}}^{\rm
  ex}}/{\partial k_x}\Big]\approx 4i\tilde{Q}^2k^2k_y\\
\hspace{-0.5cm}&&\times\left(\begin{array}{cccc}
(\gamma^2+1) & 0 & 0 & -2\gamma\\
0 & -(\gamma^2+1) & 2\gamma  & 0\\
0&2\gamma   & -(\gamma^2+1)  &0\\
-2\gamma& 0  & 0  & (\gamma^2+1)\\
\end{array}\right).
\end{eqnarray}
It is noted that here when the screening effect is considered,  $\tilde{Q}(k)$
has been treated as
  a constant.   
Accordingly, the valley Hall conductivity in BL MoS$_2$ is written as
\begin{equation}
\tilde{\sigma}_x^{z}=-2(\gamma^2+1)\frac{\displaystyle |e|}{\displaystyle h}\int_0^{\infty} d\varepsilon_k 
\frac{\displaystyle dB_k^0}{\displaystyle d\varepsilon_k}\frac{\displaystyle 4\tilde{\Omega}_k^2\tau^2}{\displaystyle
  1+4\tilde{\Omega}_k^2\tau^2},
\label{sigma_BL2}
\end{equation}
where $\tilde{\Omega}_k=\tilde{\varepsilon}_k^{\rm ex}/\hbar$. Obviously, for
the valley Hall conductivity in 
BL MoS$_2$, it is interesting to see that Eq.~(\ref{sigma_BL2}) is similar to
  Eq.~(\ref{sigma_ML2}) in ML MoS$_2$.
 Therefore, in the weak and strong scattering regimes, 
similar features for the valley Hall conductivity 
to the one in ML situation can be obtained, as addressed in Sec.~\ref{Valley_Hall_ML}.

\section{Summary}
\label{summary}
In summary, we have investigated the valley
  depolarization dynamics and valley Hall effect of exciton in ML and BL
  MoS$_2$ by solving the KSBEs.\cite{wu-review} The effect of the
  exchange-interation-modified energy spectra is explicitly
  considered.
 For the valley depolarization dynamics, in ML MoS$_2$, it is
 interesting to find that the
  conventional motional narrowing relation $\tau_s\propto \tau_k^{-1}$ in the
  strong scattering regime is no longer valid. It is
  revealed that in this regime, a novel valley
  depolarization channel is opened, where the valley lifetime first decreases and
  then increases with the increase of the disorder strength, showing
  the EY-like\cite{Yafet,Elliott} behavior from the point view of the spin
  relaxation.\cite{Awschalom,Zutic,fabian565,Dyakonov,Korn}
 This channel comes from the newly module-dependent
inhomogeneous broadening in the exciton-disorder scattering,
in which the same energy corresponds to different momentum
modules due to the exchange-interaction-modified
energy spectra. This is very different from the conventional
situation, in which the inhomogeneous broadening
comes from the angular anisotropy of the momentum in
the exciton-disorder scattering.\cite{Sham1,Sham2} Moreover, due to the
enhancement of the inhomogeneous broadening by this channel, EY-like
behavior arises in the MSS mechanism.

 For BL MoS$_2$, the PL depolarization dynamics with both the isotropic and
  anisotropic dielectric constants is investigated, which are
    found very different from the ML situation. With the isotropic
  dielectric constant, it is shown that with the
  exchange-interaction-modified energy spectra, the steady state
  revealed in our previous work\cite{Yu_BL} still exists. Whereas with the anisotropic dielectric
constant, the steady
state vanishes. However, it is found that when the dielectric constant 
is close to the isotropic situation, the PL polarization first
decreases fast and then slowly, indicating that the effective
depolarization time can be much longer than the
ML situation. 

For the valley Hall effect of exciton, the valley Hall conductivity
for ML and BL MoS$_2$ in both the weak and strong scattering regimes are
studied numerically and analytically. 
 We show that with the exciton equally 
pumped in the K and K' valleys, the
exciton states evolve into the equilibrium state with the valley polarization
parallel to the momentum-dependent effective magnetic field due to the exchange interaction. Then with
the drift of this equilibrium state due to the applied uniaxial strain, this
parallelism is broken and hence the
effective magnetic field can induce the momentum-dependent valley/PL
polarization, which accounts for the valley/PL current. This mechanism
is in analogy to the intrinsic
spin Hall effect of the
electron.\cite{SHE_KSBE,SHE_Glazov,SHE_Ka,SHE_MacDonald}

 Furthermore,
it is found that althougth the valley/PL
depolarization dynamics is very different between the ML and BL
situations, the valley Hall effect shows similar features
 in the momentum scattering dependence. Specifically, in the strong 
scattering regime, the valley Hall conductivity decreases
 with the increase of
the disorder strength ($\propto \tau^2$); whereas in the weak scattering regime, the valley Hall
conductivity saturates to a constant, which is proportional to the
population of exciton with $k=0$. Therefore, on one hand, in the dirty sample corresponding to the
strong scattering regime, the valley Hall effect is
markedly suppressed by the momentum scattering; on the other hand, 
in the weak scattering regime, the Bose system with no
Pauli blocking provides an ideal platform to realize large
``spin'' Hall conductivity, which can be much {\em larger} than
the one in the Fermi system, especially when the system
is close to the Bose-Einstein condensation. 

\begin{acknowledgments}

This work was supported
 by the National Natural Science Foundation of China under Grant
No. 11334014 and  61411136001, the National Basic Research Program
 of China under Grant No.
2012CB922002 and the Strategic Priority Research Program 
of the Chinese Academy of Sciences under Grant
No. XDB01000000.

\end{acknowledgments}

\begin{appendix}
\section{Energy spectra and projection matrix}
\label{AA}

In this appendix, we present the energy spectra and projection matrix for ML and BL
MoS$_2$. For the ML situation, the energy spectra read
\begin{equation}
E^{\rm ML}_{{\bf k},\pm}=\hbar^2{k}^2/(2m_{\rm
  ex})+{Q({k}){k}^2}\pm {Q({k}){k}^2}.
\end{equation}
The projection matrices are
\begin{equation}
T^{\rm ML}_{{\bf k},\pm}=\frac{1}{2k^2}\left(
\begin{array}{cc}
k^2 & \mp k^2_{+} \\
\mp k^2_{-} & k^2
\end{array} \right). 
\label{projection_ML}
\end{equation}
For the BL situation, the energy spectra are
\begin{eqnarray}
\nonumber
&&E^{\rm BL}_{{\bf k},1}=E^{\rm BL}_{{\bf k},2}=\hbar^2{k}^2/(2\tilde{m}_{\rm ex}),\\
\nonumber
&&E^{\rm BL}_{{\bf k},3}=\hbar^2{k}^2/(2\tilde{m}_{\rm
  ex})+2(1-\gamma){\tilde{Q}({k}){k}^2},\\
&&E^{\rm BL}_{{\bf k},4}=\hbar^2{k}^2/(2\tilde{m}_{\rm ex})+2(1+\gamma){\tilde{Q}({k}){k}^2}.
\end{eqnarray}
The corresponding projection matrices are given by
\begin{equation}
T^{\rm BL}_{{\bf k},1}=\frac{1}{2k^2}\left(
\begin{array}{cccc}
0 & 0 & 0 & 0 \\
0 & k^2 & 0 & k^2_{-}\\
0 & 0 & 0 & 0\\
0 & k^2_{+} & 0 & k^2
\end{array} \right), 
\end{equation}
\begin{equation}
T^{\rm BL}_{{\bf k},2}=\frac{1}{2k^2}\left(
\begin{array}{cccc}
k^2 & 0 & k^2_{+} & 0 \\
0 & 0 & 0 & 0\\
k^2_{-} & 0 & k^2 & 0\\
0 & 0 & 0 & 0
\end{array} \right), 
\end{equation}
\begin{equation}
T^{\rm BL}_{{\bf k},3}=\frac{1}{4k^2}\left(
\begin{array}{cccc}
k^2 & -k^2_{+} & -k^2_{+} & k^2 \\
-k^2_{-} & k^2 & k^2 &-k^2_{-}\\
-k^2_{-} & k^2 & k^2 &
-k^2_{-}\\
k^2 & -k^2_{+} & -k^2_{+} & k^2
\end{array} \right),
\end{equation}
and
\begin{equation}
T^{\rm BL}_{{\bf k},4}=\frac{1}{4k^2}\left(
\begin{array}{cccc}
k^2 & k^2_{+} & -k^2_{+} & -k^2 \\
k^2_{-} & k^2 & -k^2 &-k^2_{-}\\
-k^2_{-} & -k^2 & k^2 &
k^2_{-}\\
-k^2 & -k^2_{+} & k^2_{+} & k^2
\end{array} \right).
\end{equation}

\section{Role of off-diagonal elements of Eq.~(\ref{projection_ML}) on valley depolarization}
\label{BB}
\begin{figure}[ht]
  {\includegraphics[width=8cm]{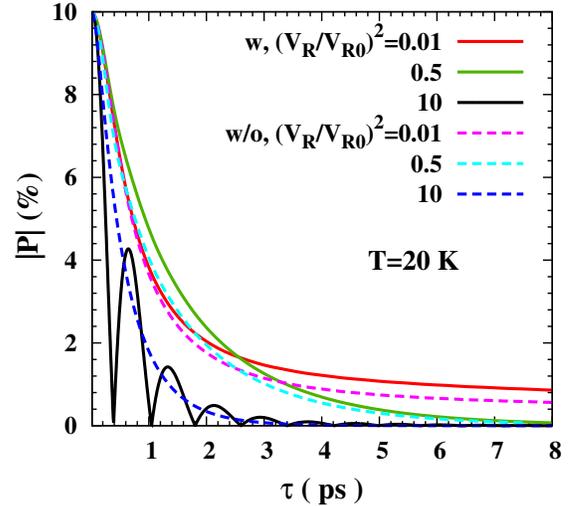}}
  \caption{(Color online) Temporal evolution of the valley
    polarization in ML MoS$_2$ with (solid curves) and without (dashed curves) the off-diagonal elements in
    Eq.~(\ref{projection_ML}). The role of the off-diagonal elements in the weak scattering
[$(V/V_{R0})^2=0.01$], EY-like [$(V/V_{R0})^2=0.5$] and
normal strong scattering [$(V/V_{R0})^2=10$] regimes are shown.}
  \label{figyw7}
\end{figure}

Here, we address the role of the 
off-diagonal elements in the projection matrix [Eq.~(\ref{projection_ML})] on
the temporal evolution of valley polarization in ML MoS$_2$. It is shown in
Fig.~\ref{figyw7} by the dashed curves that without the off-diagonal elements in
the projection matrix, the valley depolarization is a little enhanced compared
to the full calculation by the solid curves in the weak scattering
[$(V/V_{R0})^2=0.01$], EY-like [$(V/V_{R0})^2=0.5$] and
normal strong scattering [$(V/V_{R0})^2=10$] regimes. Moreover, in the normal
strong scattering regime, the off-diagonal elements of Eq.~(\ref{projection_ML})
cause the oscillations in the temporal evolution of the valley polarization, shown
as the black solid curve. By removing the off-diagonal elements in 
the projection matrix, the oscillation vanishes and the
valley polarization becomes just the oscillation exponential decay
(the blue dashed curve).

\section{Analysis on the equilibrium state}
\label{CC}
We focus on the situation where the
exchange energy is much smaller than the kinetic one. With the weak exchange
interaction $\mathcal{H}_{\bf k}^{\rm ex}$, the delta
function in Eq.~(\ref{KSBEs}) is expanded, e.g., 
\begin{eqnarray} 
\nonumber
&&\delta(E_{{\bf k'},+}-E_{{\bf k},+})=\delta({\varepsilon}_{\bf
    k'}-{\varepsilon}_{\bf k}+\varepsilon_{k'}^{\rm ex}-\varepsilon_{k}^{\rm
    ex})\\
\mbox{}&&\approx\delta({\varepsilon}_{\bf
    k'}-{\varepsilon}_{\bf k})+\frac{\partial \delta({\varepsilon}_{\bf
    k'}-{\varepsilon}_{\bf k})}{\partial \varepsilon_{\bf
    k'}}(\varepsilon_{k'}^{\rm ex}-\varepsilon_{k}^{\rm ex}).
\end{eqnarray}
With the linear order of
the exchange energy retained in the scattering term [Eq.~(\ref{KSBEs})], the KSBEs are written as  
\begin{eqnarray}
\nonumber
\hspace{-0.42cm}&&\frac{\partial \rho_{{\bf k}}}{\partial
  t}+\frac{i}{\hbar}\big[\mathcal{H}_{\bf k}^{\rm ex},\rho_{\bf
  k}\big]=\frac{2\pi}{\hbar}a^2\sum_{\bf k'}\delta(\varepsilon_{k'}-\varepsilon_{k})(\rho_{\bf k'}-\rho_{{\bf
    k}})\\
\hspace{-0.42cm}&&\mbox{}+\frac{\pi}{\hbar}a^2\sum_{\bf k'}
\frac{d\delta(\varepsilon_{k'}-\varepsilon_k)}{d \varepsilon_{k'}}\{\mathcal{H}_{\bf k}^{\rm
  ex}-\mathcal{H}_{\bf k'}^{\rm ex},\rho_{\bf k}-\rho_{\bf k'}\}.
\label{KBSE_weak}
\end{eqnarray}

In the equilibrium state, $\partial_t \rho^e_{{\bf k}}=0$ and $\big[\mathcal{H}_{\bf k}^{\rm ex},\rho^e_{\bf
  k}\big]=0$. Hence, one obtains
\begin{eqnarray}
\nonumber
\hspace{-0.36cm}&&\mbox{}\rho^e_{\bf k}
={\rho}_k^0+\int\frac{d\theta_{\bf
    k'}}{4\pi}d\delta(\varepsilon_{k'}-\varepsilon_k)\{\mathcal{H}_{\bf k}^{\rm
  ex}-\mathcal{H}_{\bf k'}^{\rm ex},\rho^e_{\bf k}-\rho^e_{\bf k'}\}.\\
\hspace{-0.36cm}
\label{iteration}
\end{eqnarray}
This integral equation can be approximately 
solved by using the iteration technique. It is assumed that $\rho^e_{\bf
  k}=B_k^0\hat{I}+\sum_{n=1}^{\infty}\rho_{\bf k}^{(n)}$.
 By substituting $B_k^0\hat{I}+\rho_{\bf k}^{(1)}$ on the left-hand side
 and $B_k^0\hat{I}$ on the right-hand side of
Eq.~(\ref{iteration}), one obtains $\rho_{\bf k}^{(1)}=\mathcal{H}_{\bf k}^{\rm
      ex}{\partial B_k^0}/{\partial \varepsilon_k}$. By repeating
    this process, one finds $\rho_{\bf k}^{(2)}$ is proportional to $(\varepsilon_k^{\rm
ex})^2$.  Here, we only keep the linear order in the exchange energy,
i.e.,
\begin{equation}
\rho_{\bf k}^e\approx B_k^0\hat{I}+\mathcal{H}_{\bf k}^{\rm
      ex}{\partial B_k^0}/{\partial \varepsilon_k}.
\end{equation}
Obviously, $\rho^e_{\bf
      k}$ commutes with $\mathcal{H}_{\bf k}^{\rm ex}$.

\section{Solution of KSBEs with an applied field}
\label{DD}
When the exchange interaction is weak, in the 
steady state ($\partial_t \rho_{{\bf k}}=0$), the KSBEs with the external force
field can be simplified to be
\begin{eqnarray}
\nonumber
\hspace{-0.48cm}&&\frac{F}{\hbar}\frac{\partial \rho_{\bf k}}{\partial {k_x}}+
\frac{i}{\hbar}\big[\mathcal{H}_{\bf k}^{\rm ex},\rho_{\bf
  k}\big]+\frac{\displaystyle
  1}{\displaystyle \tau}\big(\rho_{\bf
  k}-{\rho}_k^0\big)-\frac{1}{2\tau}\Big\{\mathcal{H}_{\bf k}^{\rm
  ex},\frac{\partial {\rho}_k^0}{\partial \varepsilon_k}\Big\}\\
\hspace{-0.48cm}&&\mbox{}
-\frac{\pi}{\tau}
\int\frac{d\varepsilon_{k'}d\theta_{\bf
    k'}}{(2\pi)^2}\frac{d\delta(\varepsilon_{k'}
-\varepsilon_k)}{d \varepsilon_{k'}}\{\mathcal{H}_{\bf k'}^{\rm
  ex},\rho_{\bf k'}\}=0.
\label{steady_E}
\end{eqnarray}
Eq.~(\ref{steady_E}) is an integral-differential equation, which can be solved
by the iteration technique approximately.
The density matrix is assumed to be 
 $\rho_{\bf k}=\rho_{\bf k}^e+\sum_{n=1}^{\infty}\rho_{\bf
    k}^{(n)}$ with $\rho_{\bf
    k}^{(n)}\propto (\varepsilon_k^{\rm ex})^n$. 

The zeroth order of Eq.~(\ref{steady_E}) is exactly Eq.~(\ref{iteration}), whose
solution has been expressed by $\rho_{\bf k}^e$ [Eq.~(\ref{steady_noE})]. The first order of Eq.~(\ref{steady_E})
reads
\begin{equation}
\rho_{\bf k}^{(1)}=
-\frac{F\tau}{\hbar}\frac{\partial \rho_{\bf k}^e}{\partial k_x}=
-\frac{F\tau}{\hbar} \frac{\partial}{\partial k_x}(B_k^0\hat{I}+\mathcal{H}_{\bf
k}^{\rm ex}\frac{\partial B_k^0}{\partial \varepsilon_k}),
\label{first_order}
\end{equation}
which is just the drift form of the equilibrium state.

 One notes that the drift density matrix $\rho^{\rm (1)}_{\bf k}$ no longer commutes
with $\mathcal{H}_{\bf k}^{\rm ex}$, which causes 
the precession of the ``spin'' vectors around the ${\bf k}$-dependent effective
magnetic field. The $n$-th order ($n\ge 2$) density matrix satisfies
\begin{eqnarray}
\nonumber
\hspace{-0.48cm}&&\frac{F}{\hbar}\frac{\partial \rho_{\bf k}^{(n-1)}}{\partial {k_x}}+
\frac{i}{\hbar}\big[\mathcal{H}_{\bf k}^{\rm ex},\rho_{\bf k}^{(n-1)}\big]
+\frac{1}{\tau}\big[\rho_{\bf k}^{(n)}-\bar{\rho}_{\bf k}^{(n-1)}\big]\\
\nonumber
\hspace{-0.48cm}&&\mbox{}-\frac{1}{2\tau}\Big\{\mathcal{H}_{\bf k}^{\rm
  ex},\frac{\partial \bar{\rho}_{\bf k}^{(n-1)}}{\partial
  \varepsilon_k}\Big\}+\frac{1}{2\tau}\int \frac{d\theta_{\bf
    k'}}{2\pi}d\delta(\varepsilon_{k'}-\varepsilon_k)\\
\hspace{-0.48cm}&&\mbox{}\times\{\mathcal{H}_{\bf k'}^{\rm
  ex},\rho_{\bf k'}^{(n-1)}\}=0,
\label{second_order}
\end{eqnarray}
where $\bar{\rho}_{\bf k}^{(n)}=1/(2\pi)\int d\theta_{\bf k}\rho_{\bf k}^{(n)}$.
Eq.~(\ref{second_order}) is complex, but fortunately it can be much simplified if only the
density matrix in the linear order of $F$ is retained (linear regime).  Furthermore, $\bar{\rho}_{\bf k}^{(n)}$
($n\ge 1$) and the last term on the left-hand side of
Eq.~(\ref{second_order}) are exactly zero due to the angle integration. Finally,
one obtains ($n\ge 2$),
\begin{equation} 
(i/\hbar)\big[\mathcal{H}_{\bf k}^{\rm ex},\rho_{\bf
    k}^{(n-1)}\big]+\rho_{\bf k}^{(n)}/\tau=0.
\end{equation}

With $\rho_{\bf k}^{(1)}$ [Eq.~(\ref{first_order})] known, $\rho_{\bf k}^{(n)}$ ($n\ge
2$) can be obtained. By summing $\rho_{\bf k}^{(n)}$, one comes to a {\it closed} form of the
density matrix for ML MoS$_2$
 (for BL MoS$_2$, one replaces $\mathcal{H}_{\bf k}^{\rm ex}$ by
 $\mathcal{\tilde{H}}_{\bf k}^{\rm ex}$, and $\Omega_k$ by $\tilde{\Omega}_k$),
\begin{eqnarray}
\nonumber
\hspace{-0.36cm}&&\rho_{\bf k}\approx 
\big(B_k^0\hat{I}+\mathcal{H}_{\bf
k}^{\rm ex}\frac{\partial B_k^0}{\partial \varepsilon_k}\big)-\frac{F}{\hbar}\tau \frac{\partial B_k^0}{\partial
k_x}\hat{I}+\frac{i}{\hbar^2}F
\frac{\partial B_k^0}{\partial \varepsilon_k}\frac{\tau^2}{1+4\Omega_k^2\tau^2}\\
\hspace{-0.36cm}&&\mbox{}\times\big[\mathcal{H}_{\bf
k}^{\rm ex},\frac{\partial \mathcal{H}_{\bf k}^{\rm
ex}}{\partial k_x}\big]
-\frac{F}{\hbar}\frac{\partial \mathcal{H}_{\bf k}^{\rm
    ex}}{\partial k_x}\frac{\partial B_k^0}{\partial \varepsilon_k}
\frac{\tau}{1+4\Omega_k^2\tau^2}.
\label{final}
\end{eqnarray}

\end{appendix}

\end{document}